\documentstyle[12pt,epsf]{article} 

\makeatletter
\def\lesssim{\mathrel{\mathpalette\vereq<}}
\def\vereq#1#2{\lower3pt\vbox{\baselineskip1.5pt \lineskip1.5pt
\ialign{$\m@th#1\hfill##\hfil$\crcr#2\crcr\sim\crcr}}}
\def\gtrsim{\mathrel{\mathpalette\vereq>}}
\def\alt{\lesssim}
\def\agt{\gtrsim}
\makeatother

\title{
{\normalsize E-print hep-ph/9903401 \hfill Preprint YARU-HE-98/11} \\[7mm]
{\LARGE\bf 
       Axion Decay of a Photon in an External Electromagnetic Field}}

\author{\Large N.V.~Mikheev, A.Ya.~Parkhomenko and L.A.~Vassilevskaya\\
       {\it Yaroslavl (Demidov) State University,} \\ 
       {\it Sovietskaya 14, Yaroslavl 150000, Russia}}

\date{}

\begin{document}

\maketitle

\begin{abstract}
An interaction of a pseudoscalar particle with two photons induced
by an external electromagnetic field is used to study the photon decay  
$\gamma \to \gamma a$ where $a$ is a pseudoscalar particle associated 
with the Peccei-Quinn $U(1)$ symmetry. 
The field-induced axion emission by photon is analyzed as a 
possible source of energy losses by astrophysical objects. 
\end{abstract}


\section{Introduction}

The physical nature of dark matter in the universe still remains
an unresolved mystery. Axions~\cite{Peccei77,WW} which appear as
Nambu-Goldstone bosons of the spontaneously broken Peccei-Quinn 
symmetry $U_{PQ}(1)$ are one of the well-motivated candidates for
the cold dark matter~\cite{Raffelt-book,Raffelt-castle97,Raffelt-school97}.
In analogy to neutral pions, axions generically interact with
photons according to Lagrangian: 
\begin{equation}
{\cal L}_{a \gamma} = -\frac{g_{a\gamma}}{4} F_{\mu \nu} 
\tilde F^{\mu\nu} a =
g_{a\gamma} \;{\bf E} \cdot {\bf B} \; a
\label{eq:L-ag}
\end{equation}
\noindent
with a strength 
\begin{equation}
g_{a\gamma} = \frac{\alpha}{\pi f_a} \, \xi, 
\label{eq:ag-couple}
\end{equation}
\noindent
where $f_a$ is the energy scale of the symmetry breaking and $\xi$ 
is a model-dependent factor of order unity, $F_{\mu\nu}$ is the
electromagnetic field strength tensor, $\tilde F_{\mu\nu}$ its dual, 
and {\bf E} and {\bf B} are the electric and magnetic fields, 
respectively; $a$ is the axion field. Existing axion models also contain 
an interaction of axions with charged fermions (usual or exotic):
\begin{equation}
{\cal L}_{af} =  \frac{ g_{af}}{2 m_f} 
(\bar f \gamma_\mu \gamma_5 f) \, \partial_\mu \, a,
\label{eq:L-af}
\end{equation}

\noindent which automatically leads to an electromagnetic coupling 
of the form in Eq.~(\ref{eq:L-ag}) because of the triangle loop  
$a \gamma \gamma$--amplitude~\cite{Raffelt-book}. Here 
$g_{af} = C_f m_f/f_a$ is a dimensionless coupling constant, 
$C_f$ is a model--dependent factor, $f$ is the fermion field with 
mass $m_f$; $\gamma_\mu$, $\gamma_5$ are Dirac $\gamma$-matrices.

The two-photon-axion interaction vertex allows for the axion 
radiative decay $a \to \gamma\gamma$~\cite{Raffelt-book},
for the Primakoff conversion $a \leftrightarrow \gamma$
in the presence of electric or magnetic fields~\cite{Dicus} 
as well as for plasmon decay 
$\gamma_T \to \gamma_L a$~\cite{Pantziris,Raffelt-88} 
and coalescence $\gamma_L \gamma_T \to a$~\cite{Raffelt-88}. 
The last two processes are kinematically possible 
because of the dispersion relations of electromagnetic excitations  
in a plasma which differ significantly from the vacuum dispersion.

It is known that an external electromagnetic field plays the
role of an active medium in its influence on particle properties.
Due to a nontrivial kinematics of charged particles and 
a photon dispersion in the external field, such axion processes as
the axion decay into electron-positron pair 
$a \to e^+ e^-$~\cite{MVaff-plb,MOVaff-mpla},
the axion cyclotron emission $e \to e a$~\cite{cyclotron,MRV} are not 
only opened kinematically but also become substantial.

Due to a very weak interaction of axions with a matter (the latest 
astrophysical data yield $f_a \agt 10^{10}$~GeV~\cite{Raffelt-book}), 
the processes with axion emission could be 
of great importance in astrophysics as an additional
source of star energy losses~\footnote{For example, the upper bound 
on $m_a$ is obtained from the requirement that stars not lose too much 
energy by axions.}. In the studies of processes occurring inside 
astrophysical objects, one has to take into account the influence 
of both components of the active medium -- a plasma and a magnetic field. 
A situation is also possible when the field component dominates; 
for example, in a supernova explosion or in a coalescence of neutron 
stars, a region outside the neutrino sphere of order of hundred
kilometers with a strong magnetic field and 
a rather rarefied plasma could exist. 

In this letter we investigate a forbidden in vacuum axion 
decay of photon $\gamma \to \gamma a$ in an external electromagnetic 
field using the effective $a \gamma \gamma$-interaction~\cite{MPV} and 
estimate a possible influence of this decay on the star cooling process.

\section{Field-Induced ``Effective Masses'' of the Particles}

In calculating $\gamma \to \gamma a$ decay in the external field, 
we have to integrate over the phase space of the final particles 
(photon, axion)
taking into account their non-trivial kinematics. The kinematics
depend substantially on the particles dispersion relations in the
electromagnetic field which plays the role of an anisotropic medium. 
The photon ``effective masses'' squared $\mu^2_\lambda$ induced by
the external field are defined as the eigenvalues 
of the photon polarization operator $\Pi_{\mu \nu}$:
\begin{eqnarray}
\Pi_{\mu \nu} & = & i \, \sum_{\lambda=1}^3 \Pi^{(\lambda)} \, 
\varepsilon^{(\lambda)}_\mu \;\varepsilon^{(\lambda)}_\nu ,
\label{eq:PO} 
\end{eqnarray}
where $\varepsilon^{(\lambda)}_\mu$ are the photon polarization
vectors. Note that eigenmodes of two transverse photons with 
polarization vectors $\varepsilon^{(1)}_\mu$ and $\varepsilon^{(2)}_\mu$: 
\begin{eqnarray}
\varepsilon^{(1)}_\mu & = & \frac{(q F)_\mu}{\sqrt{(qFFq)}} , 
\quad 
\varepsilon^{(2)}_\mu = \frac{(q \tilde F)_\mu}{\sqrt{(qFFq)}} , 
\label{eq:polar} 
\end{eqnarray}
are realized in the electromagnetic field. Here $F_{\mu\nu}$ is the 
external field tensor and $q_\mu$ is the photon four-momentum. 

Note that an arbitrary weak smooth field in which a relativistic particle 
propagates is well-described by the constant crossed field limit 
(${\bf E} \perp {\bf B}$, $E = B$). Thus, calculations in this field 
possess a great extent of generality and acquire interest by themselves. 
Below we will consider the constant crossed field as the external field.
In this case the dynamic parameter $\chi$ is the only field invariant. 

\begin{figure}[tb]
\centerline{\epsfxsize=.4\textwidth \epsffile[225 320 425 610]{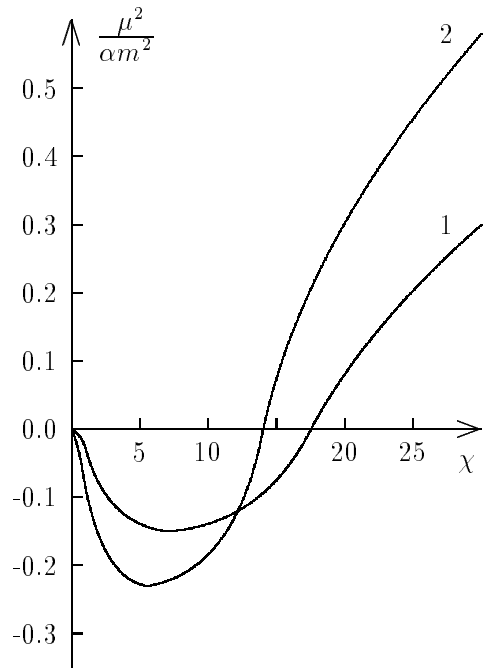}}
\caption{The photon dispersion curves in the crossed field.}
\label{fig:disp}
\end{figure}

The analysis of $\Pi_{\mu \nu}$ in one-loop approximation in the crossed 
field~\cite{Ritus} shows that the dispersion curves (Fig.~\ref{fig:disp})
corresponding to the above-mentioned photon eigenmodes~(\ref{eq:polar}),
though being similar in their qualitative behavior, are quantitatively 
different. The difference in values of the field-induced  
``effective masses'' squared of the first- and second-photon eigenmodes 
makes the process of the photon splitting $\gamma \to \gamma a$ possible, 
where $a$ is an arbitrary relatively light pseudoscalar with mass $m_a$.

The field-induced contribution to the small axion 
mass~\footnote{The allowed range for the axion mass is strongly 
constrained by astrophysical and cosmological considerations, 
as a result $10^{-5} \, \mbox{eV} \alt m_a \alt 10^{-2} \, 
\mbox{eV}$~\cite{Raffelt-book,Raffelt-castle97,Raffelt-school97}.}
$\delta m_a$ (the real part of $a \to f \tilde f \to a$ transition 
amplitude via the fermion loop) is negligible:
\begin{equation}
\frac{\delta m_a^2}{m_a^2} \alt 10^{-10} \, C_e^2 \, \chi^{2/3}, 
\label{eq:ax-mass}
\end{equation}
where $C_e$ is the model-dependent factor which determines the 
axion-electron coupling constant $g_{ae}$. 
Hereafter we will take axion as a massless particle. 

\section{Matrix Element}

The effective axion-photon vertex was obtained in Ref.~\cite{MPV} 
and used to analyze the axion radiative decay 
$a \to \gamma\gamma$~\cite{MV-agg}. As in the case of $a \to \gamma\gamma$, 
the main contribution to the amplitude of the photon decay
$\gamma \to \gamma a$ comes from the bilinear on the external field terms
of $a \gamma \gamma$-vertex and can be presented in the form:
\begin{eqnarray}
& & M \simeq \frac{\alpha}{\pi} \sum_f \frac{Q^2_f g_{af}}{m_f}
\left \{ \frac{\chi}{\chi_2} (f_1 \tilde {\cal F}) (f_2 {\cal F}) 
J(\chi_1,\chi_2)
- \frac{\chi}{\chi_1} (f_2 \tilde {\cal F}) (f_1 {\cal F}) 
J(\chi_2,\chi_1)
\right \} ,
\label{eq:Ampl} 
\end{eqnarray}
\begin{eqnarray}
{\cal F}_{\mu\nu} & = & \frac{F_{\mu\nu}}{B_f}, \,\,\,\,
{\tilde{\cal F}_{\mu \nu}} = \frac{1}{2} \,
\varepsilon_{\mu \nu \alpha \beta}
{\cal F}_{\alpha \beta},  \,\,\,\,
B_f  =  \frac{m^2_f}{e Q_f},
\nonumber \\
(f_i)_{\alpha \beta} & = & q_{i\alpha} \varepsilon_{i\beta} -
q_{i\beta} \varepsilon_{i\alpha} ,
\,\,
(\tilde f_i)_{\alpha \beta} = \frac{1}{2} \,
\varepsilon_{\alpha \beta \mu \nu} \, (f_i)_{\mu \nu} ,
\,\, i = 1, 2,
\nonumber 
\end{eqnarray}
\begin{eqnarray}
J(\chi_1, \chi_2) & = & \int\limits_0^1 dx \int \limits_0^{1-x}
dy \, y \, (1 - y - 2x) \, \eta^3 \; (1 - \eta \, f(\eta)),
\label{eq:IntJ} \\
f(\eta) & = & i \int\limits_0^\infty du \exp \left [ - i \left ( \eta u +
{1\over 3} u^3 \right ) \right ],
\nonumber \\
\eta(\chi_1,\chi_2) & = & \Big [ \left ( x (1 - x) \chi_2 
+ y (1 - y) \chi_1 \right )^2
\nonumber \\
     & - & 4 x y \left ( x y + (1 - x - y )(x + y) \right ) \chi_1 \chi_2
\Big ]^{-1/3},
\nonumber 
\end{eqnarray}
where $f(\eta)$ is the Hardy-Stokes function, $\chi$, $\chi_1$
and $\chi_2$ are the dynamic parameters
\begin{eqnarray}
\chi^2 =  \frac{(p {\cal F} {\cal F} p)}{m_f^2}, \qquad
\chi_i^2 = \frac{(q_i {\cal F} {\cal F} q_i)}{m_f^2},\;\;\; i = 1,2.
\label{eq:Chi} 
\end{eqnarray}
Here $q_1$ and $q_2$ are the four-momenta of the initial and final 
photons, respectively; $p$ is the axion four-momentum. The summation in
Eq.~(\ref{eq:Ampl}) is over the virtual fermions $f$ with mass $m_f$ 
and the relative electric charge $Q_f$; $e > 0$ is the elementary 
charge; $B_f$ is the critical value of the magnetic field strength for 
the fermion. We have neglected terms of order $m_a^2/E_a^2$, 
$\mu_i^2/E_a^2$ in the argument of the function $f(\eta)$.

The amplitude, Eq.~(\ref{eq:Ampl}), is significantly simplified in the
case of small values of the dynamic parameters. In this case the decay 
of the photon with the first polarization (see Fig.~\ref{fig:disp}) is 
allowed kinematically due to the condition $\mu^2_1 > \mu^2_2$:
\begin{eqnarray}
\gamma^{(1)} \rightarrow \gamma^{(2)} + a.
\label{eq:G1}
\end{eqnarray}
With the photon polarization vectors, Eq.~(\ref{eq:polar}), 
the amplitude is:
\begin{eqnarray}
 M  & \simeq & - \frac{4 \alpha}{\pi} t\;(1 - t) \sum_f \;
Q^2_f g_{af} m_f \; \chi_1^2 \; J(t \chi_1, \chi_1) ,
\label{eq:M5}
\end{eqnarray}
where $t = \omega_2/\omega_1$ is the relative energy of the
final photon. The function $J$ defined in~(\ref{eq:IntJ}) has the
following asymptotic behavior at small values of its arguments:
\begin{eqnarray}
J(t \chi_1, \chi_1) \bigg \vert_{\chi_1 \ll 1}  \simeq \frac{2}{63} \;
\chi_1^2 \; (1 - 2t) + O(\chi_1^4).
\nonumber
\end{eqnarray}

The limit of large values of the dynamic parameter can only be 
realized in the central regions of astrophysical objects with 
the strong magnetic field where the characteristic temperatures
$T \agt 10 \, \mbox{MeV}$. But in these regions from both components 
of the active medium, the plasma and the magnetic field, the plasma 
component dominates. Thus Eq.~(\ref{eq:Ampl}) cannot be used to 
estimate possible astrophysical applications
because it does not take into account the plasma influence.  

\section{Axion Emissivity}

The probability of the photon decay in the limit of small values
of dynamic parameter $\chi_1$ can be presented in the form:
\begin{eqnarray}
W^{(F)} = {1 \over 16\pi \omega_1}
\int \limits_{0}^1 d t \, \vert  M \vert^2 
\simeq 
4.8 \cdot 10^{-6} \,
\left( {\alpha \over \pi}\right )^2
\frac {\left( \sum_f Q^2_f g_{af} m_f \chi_1^4  \right )^2}
{\pi \omega_1}.
\label{eq:WFll1} 
\end{eqnarray}

To illustrate a possible application of the result obtained, we calculate 
the contribution of this process to the axion emissivity $Q_a$, 
i.e. the rate of energy loss per unit volume, of the photon gas: 
\begin{eqnarray}
Q_a = \int \; \frac{d^3 q_1}{(2 \pi)^3} \; 
\omega_1 \, n_B (\omega_1) \;
\int \limits_0^1 d t \, \frac{d W^{(F)}}{d t}\; (1 - t) \;
(1 + n_B (\omega_1 t)),
\label{eq:Q1} 
\end{eqnarray}
where $n_B(\omega_1)$ and $n_B(\omega_1 t)$ are the Planck distribution
functions of the initial and final photons at temperature $T$, respectively.
In Eq.~(\ref{eq:Q1}) we have taken into account that the photon
of only one polarization (the first one in our case) splits.
Note that the dynamic parameter $\chi_1$ in $d W^{(F)}/d t$
depends on the photon energy $\omega_1$ and the angle $\theta$ between
the initial photon momentum ${\bf q_1}$ and the magnetic field
strength {\bf B}:
\begin{eqnarray}
\chi_1 = \frac{\omega_1}{m_f}\; \frac{B}{B_f} \; \sin \theta.
\nonumber
\end{eqnarray}
Neglecting the photon ``effective masses'' squared
($\omega_1^2 = | {\bf q_1} |^2 + \mu_1^2 \simeq | {\bf q_1} |^2$),
the result of the calculation of the axion emissivity
$Q_a$~(\ref{eq:Q1}) becomes:
\begin{eqnarray}
Q_a & \simeq & 2.15 \;\frac{\alpha^2}{\pi^5}\; \left (
\sum_f \frac{Q_f^2 \; g_{af}}{m_f^3 \; B_f^4} \right )^2 \;
T^{11} \; B^8 .
\label{eq:Q2}
\end{eqnarray}

Below we estimate the contribution of the photon decay
$\gamma \to \gamma a$ into the axion luminosity $L_a$ in a supernova 
explosion from a region of order of hundred kilometers in size
outside the neutrino sphere. In this region a rather rarefied plasma
with the temperature of order of MeV and the magnetic field of order
$10^{13}$~G can exist. Under these conditions the estimation of the
axion luminosity is:
\begin{eqnarray}
L_a \simeq 3 \times 10^{33} \; \frac{\mbox{erg}}{\mbox{s}} 
\left ( \frac{g_{ae}}{10^{-13}} \right )^2 
\left ( \frac{T}{1 \;\mbox{MeV}} \right )^{11} 
\left ( \frac{B}{10^{13}\; \mbox{G}} \right )^8 
\left ( \frac{R}{100 \;\mbox{km}}\right )^3 .
\label{eq:Lum} 
\end{eqnarray}
The comparison of Eq.~(\ref{eq:Lum}) with the total neutrino luminosity
$L_\nu \sim 10^{52}$~erg/s from the neutrino sphere shows that the
contribution of the photon decay to the supernova energy losses 
is very small and does not allow  to get a new restriction on the
axion-electron coupling $g_{ae}$.

\section{Conclusions}

In this letter we have studied the field-induced
photon decay $\gamma \to \gamma a$ ($a$ is a light 
pseudoscalar particle). For the pseudoscalar particle, we considered 
the most widely discussed particle, the axion, corresponding to the 
spontaneous breaking of the Peccei-Quinn symmetry.
This is forbidden in vacuum process but it becomes kinematically possible
because photons of different polarizations obtain different 
field-induced ``effective masses'' squared. At the same time an 
external field influence on the axion mass is negligible. 

The process $\gamma \to \gamma a$ could be of interest as an additional 
source of energy losses by astrophysical objects. We considered 
the case of small values of the dynamic parameter which can be 
realized, for example, in a supernova explosion. 
While this process and its evaluation are conceptually quite 
intriguing, the actual energy-loss rate appeared to be rather small in
comparison with the neutrino luminosity in the conditions considered.

\section*{Acknowledgments} 

This work was partially supported by INTAS under grant No.~96-0659 
and by the Russian Foundation for Basic Research 
under grant No.~98-02-16694.
The work of N.V.~Mikheev was supported under grant No.~d98-181
by International Soros Science Education Program.


\begin{thebibliography}{15}
%
\bibitem{Peccei77}
   R.D.~Peccei and H.R.~Quinn,
   {\it Phys.~Rev.~Lett.} {\bf 38}, 1440 (1977);
   {\it Phys. Rev.} {\bf D16}, 1791 (1977).
%
\bibitem{WW}
   S.~Weinberg,
   {\it Phys.~Rev.~Lett.} {\bf 40}, 223 (1978); 
   F.~Wilczek,
   {\it ibid.} {\bf 40}, 279 (1978).
%
\bibitem{Raffelt-book}
   G.G.~Raffelt, 
   {\it Stars as Laboratories for Fundamental Physics} 
   (University of Chicago Press, 1996).
%
\bibitem{Raffelt-castle97}
   G.G.~Raffelt, 
   in {\it Proceedings of Beyond the Desert 1997}, 
   eds. H.V.~Klapder-Kleingrothaus and H.~Paes. 
   (Institute of Physics Pub., 1998), p.~808; 
   preprint astro-ph/9707268.
%
\bibitem{Raffelt-school97}
   G.G.~Raffelt, 
   in {\it Proceedings of 1997 European School of High-Energy Physics}, 
   eds. N.~Ellis and M.~Neubert. (CERN, Geneva, 1998), p.~235;  
   preprint hep-ph/9712538.
%
\bibitem{Dicus}
  D.A.~Dicus {\it et al.}, 
  {\it Phys. Rev.} {\bf D18}, 1829 (1978).
%
\bibitem{Pantziris}
  A.~Pantziris and K.~Kang, 
  {\it Phys. Rev.} {\bf D33}, 3509 (1986).
%
\bibitem{Raffelt-88}
  G.G.~Raffelt,
  {\it Phys. Rev.} {\bf D37}, 1356 (1988).
%
\bibitem{MVaff-plb}
   N.V.~Mikheev and L.A.~Vassilevskaya, 
   {\it Phys. Lett.} {\bf B410}, 203 (1997).
%
\bibitem{MOVaff-mpla}
   N.V.~Mikheev, O.S.~Ovchinnikov and L.A.~Vassilevskaya, 
   {\it Mod. Phys. Lett.} {\bf A13}, 321 (1998).
%
\bibitem{cyclotron}
   A.V.~Borisov and V.Yu.~Grishina, 
   {\it JETP} {\bf 79}, 837 (1994);
   M.~Kachelriess, C.~Wilke and G.~Wunner, 
   {\it Phys. Rev.} {\bf D56}, 1313 (1997).
%
\bibitem{MRV}
   N.V.~Mikheev, G.G.~Raffelt and L.A.~Vassilevskaya,
   {\it Phys. Rev.} {\bf D58}, 055008 (1998);  
   preprint hep-ph/9803486.
%
\bibitem{MPV}
   L.A.~Vassilevskaya, N.V.~Mikheev and A.Ya.~Par\-kho\-men\-ko, 
   {\it Phys. At. Nucl.} {\bf 60}, 2041 (1997).
%
\bibitem{Ritus}
   V.I.~Ritus, 
   in {\it Issues in Intense-Field Quantum Electrodynamics} 
   (Nauka, 1986), p.~141.  
%
\bibitem{MV-agg}
   N.V. Mikheev and L.A. Vassilevskaya, 
   {\it Phys. Lett.} {\bf B410}, 207 (1997).
%
\end{thebibliography}
\end{document}